\documentclass[a4paper,11pt]{article}
\pdfoutput=1 
\usepackage{jcappub}
\usepackage[T1]{fontenc} 
\makeatletter
\def\@fpheader{\relax}
\makeatother
\usepackage{xcolor}

\usepackage{subcaption}

\title{\boldmath Constraining Gamma-ray Lines from Dark Matter Annihilation using Fermi-LAT and H.E.S.S. data}

\author[a,b,c]{Lucia Angel,}
\author[d]{Guillermo Gambini,}
\author[a,b,c]{Leticia Guedes,}
\author[a,b,c]{Farinaldo S. Queiroz,}
\author[d]{Vitor de Souza}

\affiliation[a]{Departamento de F\'isica, Universidade Federal do Rio Grande do Norte, 59078-970, Natal, RN, Brasil}
\affiliation[b]{International Institute of Physics, Federal University of Rio Grande do Norte, Campus Universit\'ario, Lagoa Nova, Natal-RN 59078-970, Brazil}  
\affiliation[c]{Millennium Institute for Subatomic Physics at the High-Energy Frontier (SAPHIR) of ANID, Fern\'andez Concha 700, Santiago, Chile}
\affiliation[d]{Instituto de Física de São Carlos, Universidade de São Paulo, Av. Trabalhador São Carlense 400, São Carlos-SP, 13566-590, Brazil}

\emailAdd{lucia.correa.717@ufrn.edu.br}
\emailAdd{guillermo.gambini@ifsc.usp.br}
\emailAdd{leticia.guedes.110@ufrn.edu.br}
\emailAdd{farinaldo.queiroz@ufrn.br}
\emailAdd{vitor@ifsc.usp.br}

\abstract{Using 14 years of Fermi-LAT data and 10 years of H.E.S.S. observations in the direction of the galactic center, we derive limits on gamma-ray lines originated from dark matter annihilations for fermionic and scalar fields. We describe the dark matter annihilation into $\gamma \gamma$ or $\gamma Z$ final states in terms of effective operators and place limits on the energy scale as a function of the dark matter mass taking into account the energy resolution of the instruments. For the Fermi-LAT data, we considered an NFW and a contracted NFW dark matter density profile, the latter being preferred by the Fermi GeV excess. For the H.E.S.S. observation, we used an NFW and Einasto profile. Fermi-LAT yields the most stringent constraints for dark matter masses below 300 GeV, whereas H.E.S.S. has the strongest ones for dark matter masses above 1 TeV. The telescopes share similar sensitivities for dark matter masses between 300 GeV and 1 TeV. We conclude that Fermi-LAT (H.E.S.S.) can probe energy scales up to $10(20)$~TeV for scalar and fermionic dark matter particles.}

\begin{document} 
\maketitle
\flushbottom

\section{Introduction}
\label{sec:intro}

The nature of dark matter has been established from several cosmological and astronomical observations, yet its nature remains unknown \cite{Bertone:2004pz}. Numerous experimental endeavors are underway to detect non-gravitational interactions of dark matter, employing methods such as the direct detection of its scattering off nuclei, and direct production in particle accelerators. An orthogonal search strategy involves the detection of the byproducts of dark matter annihilation into Standard Model (SM) particles \cite{Arcadi:2017kky}. A particularly noteworthy aspect of this search is the transparency of the Universe to gamma rays with energies around $100$~GeV over galactic distances, facilitating the use of their spatial distribution and energy spectrum as tools to distinguish potential dark matter signals from often ambiguous astrophysical backgrounds \cite{Cirelli:2010xx, HESS:2011zpk, Bringmann:2012ez}.

As it is often assumed that dark matter particles interacted with SM particles in the early universe, it is reasonable to assume that such interactions might still take place today, especially in astrophysical regions which are dark matter dominated such as dwarf spheroidal galaxies. In this aspect, dark matter annihilations into SM particles will yield a continuous gamma-ray emission which is not easily distinguished from other astrophysical emissions \cite{Abazajian:2010zb,Abazajian:2010sq,Bringmann:2011ye,Abazajian:2011tk,Abazajian:2011ak,Fermi-LAT:2015att,Baring:2015sza,Garcia-Cely:2013zga,Queiroz:2016zwd,Profumo:2016idl,HESS:2016mib,Profumo:2017obk,Queiroz:2019acr,Abazajian:2020tww,Siqueira:2021lqj}. 
Nevertheless, dark matter annihilation into photon pairs produces monochromatic photons, which represent a smoking-gun signature of dark matter annihilation due to a relative lack of confounding astrophysical backgrounds. An eventual observation of gamma-ray lines will point to the dark matter mass, as $E_\gamma=m_{\chi}$, where $m_{\chi}$ is the dark matter mass. Thus, the search for gamma-ray lines from dark matter annihilation is worth pursuing.  That said, H.ES.S. collaboration searched for gamma-ray lines in the 300 GeV - 70 TeV energy range, in the central region of the Milky Way halo which is one of the most promising targets due to the large dark matter density and its proximity to Earth. H.E.S.S used a two-dimensional maximum likelihood method to account for the spectral and spatial features of the dark matter signal and have a better discrimination power over the background. No signal gamma-ray excess has been found during the ten years of observations which led to an upper bound on the dark matter annihilation cross-section into gamma-ray lines of $4\times 10^{ -28}cm^3s^{-1}$ for $m_{\chi}=1$ TeV \cite{HESS:2018cbt} assuming the dark matter density to be distributed in the galaxy obeying an Einasto profile \cite{Graham:2005xx,Navarro:2008kc}. Fermi-LAT telescope has collected over a decade of data in the direction of the galactic center  \cite{Fermi-LAT:2015sau}. Fermi-LAT collaboration has reported a search for the first 62 months of the mission in the gamma-ray energy range of $1-100$~GeV from a $15^o$ degree region centered in the GC. An updated study includes fourteen years of data with photons from $10$~GeV to $2$~TeV, the new event selection (PASS-8), and a much larger region consisting of the inner $30^o$ of our galaxy \cite{Foster:2022nva}. An upper bound on the dark matter annihilation cross-section into gamma-ray lines of $6 \times 10^{-30} cm^3 s^{-1}$ was found adopted a Navarro-Frenk-White profile \cite{Navarro:1995iw} and a contracted density profile which is preferred by the dark matter interpretation of the Fermi GeV gamma-ray excess \cite{Daylan:2014rsa}. We will use these findings to place a limit on the fermionic and scalar dark matter models that annihilate into gamma-ray lines using effective operators. A significant improvement is expected once the Cherenkov Telescope Array starts taking data \cite{Silverwood:2014yza,Pierre:2014tra,Lefranc:2015pza,Queiroz:2015utg,Garcia-Cely:2015khw,Ibarra:2015tya,CTAConsortium:2017dvg,CTA:2020qlo,NFortes:2022dkj,CherenkovTelescopeArray:2023aqu,CTAConsortium:2023yak}. In this study, we focus on real data and not future projections, and for this reason, we rely on Fermi-LAT and H.E.S.S. data which yield the most relevant constraints in the mass range compared to other instruments \cite{MAGIC:2016xys,e-ASTROGAM:2017pxr,Acharyya:2023ptu}.

Therefore, our work extends previous works \cite{Rajaraman:2012db} that addressed dark matter annihilation into gamma-ray lines in the context of effective operators because we use an updated data set from both Fermi-LAT and H.E.S.S. telescope, and we consider both fermionic and scalar dark matter concluding that those searches for gamma-ray lines can probe new physics energy scales well above the $10$~TeV. 
\begin{center}
\begin{table}[h]
\begin{tabular}{|c|c|c|}
\hline
\multicolumn{3}{|c|}{Scalar Dark Matter Operators - Dimension Six} \\
\hline
S1 & $\frac{1}{\Lambda^2_{S1}}\chi\chi^* B_{\mu\nu}B^{\mu\nu}$ & $\gamma \gamma$ \\
\hline
S2 & $\frac{1}{\Lambda^2_{S2}}\chi\chi^* W^a_{\mu\nu}W^{a\mu\nu}$ & $\gamma \gamma$, $\gamma Z$ \\
\hline
S3 & $\frac{1}{\Lambda^2_{S3}}\chi\chi^* B_{\mu\nu}\tilde{B}^{\mu\nu}$ & $\gamma \gamma$\\
\hline
S4 & $\frac{1}{\Lambda^2_{S4}}\chi\chi^* W^a_{\mu\nu}\tilde{W}^{a\mu\nu}$ & $\gamma \gamma$, $\gamma Z$ \\
\hline
\end{tabular}
\caption{List of effective interactions for complex scalar dark matter and the type of line signals ($\gamma \gamma$,
$\gamma Z$).}
\label{Operatorlist-scalar}
\end{table}
\end{center}
\begin{center}
\begin{table}[h]
\begin{tabular}{|c|c|c|}
\hline
\multicolumn{3}{|c|}{Fermion Dark Matter - Dimension 5 Operators} \\
\hline
F1 & $\frac{1}{\Lambda_{A1}}\bar{\chi}\gamma^{\mu\nu}\chi B_{\mu\nu}$ &  $\gamma \gamma$ \\
\hline
F2 & $\frac{1}{\Lambda_{A2}}\bar{\chi}\gamma^{\mu\nu}\chi \tilde{B}_{\mu\nu}$ &  $\gamma \gamma$, $\gamma Z$ \\
\hline
\end{tabular}
\caption{List of effective interactions for Dirac fermion dark matter and the type of line signals ($\gamma \gamma$,
$\gamma Z$).}
\label{Operatorlist-fermion}
\end{table}
\end{center}
\section{Dark Matter Signal}

\begin{figure}[h]
  \centering
  \includegraphics[scale=0.5]{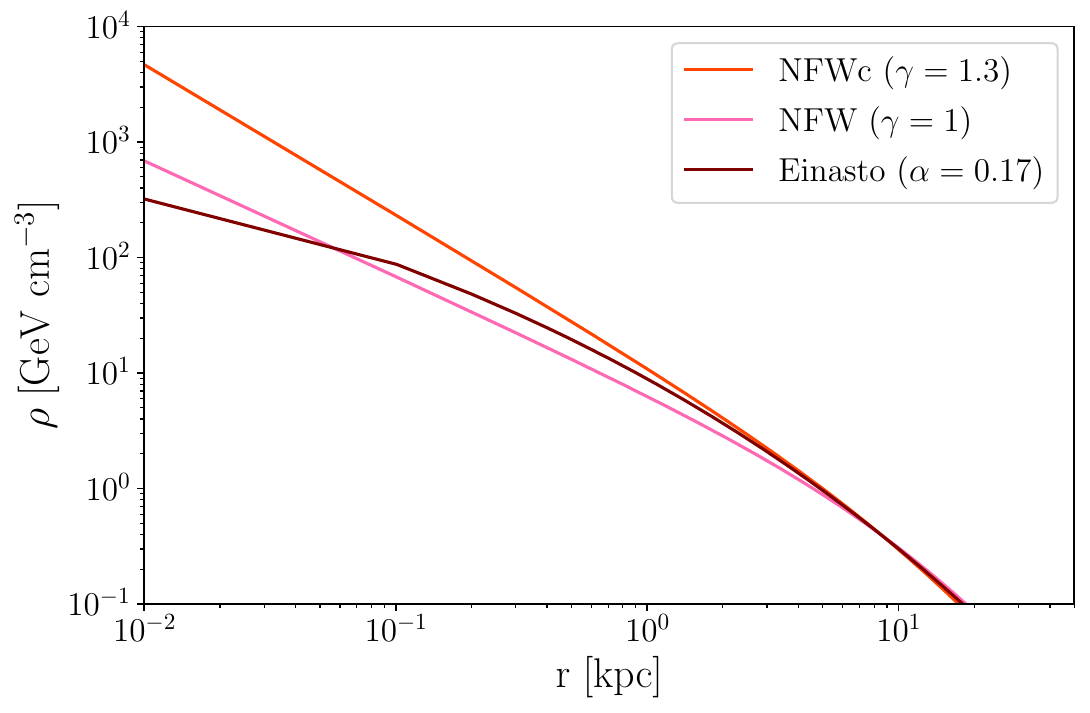}
  \caption{Comparison between NFW, NFW contracted (NFWc) and Einasto dark matter density profiles along the distance}
  \label{fig:prof}
\end{figure}

The flux of gamma-rays with energy $E = m_\chi$ from the $\chi \chi \to \gamma\gamma$ annihilation process in a given solid angle element of the sky ($d\Omega$) is given by \cite{Gaskins:2016cha},
\begin{equation}
   \frac{d \Phi}{d \Omega} 
=   \frac{\langle \sigma v \rangle}{4 \pi m_\chi^2} 
     \,  J_\Omega \, \mathcal{E}(E)\,,
     \label{eqDMflux}
\end{equation}where $\langle \sigma v \rangle$ is the velocity averaged annihilation cross-section, $\mathcal{E}$ is effective area times exposure time, and $J_\Omega$ the astrophysical factor defined to be,
\begin{equation}
    \mathcal{J}(\Omega) = \int_{l.o.s} ds \rho^2 \left(\vec{r}(s,\Omega)\right)\,,
\end{equation}with $\rho_(\vec{r})$ being dark matter density along the line-of-sight (l.o.s), located at a distance $s$ to the Earth. Using the dark matter signal given in Eq.\eqref{eqDMflux}, a bound on the dark matter annihilation cross section into gamma-ray lines was derived by Fermi-LAT collaboration using $\sim$~6 years of data \cite{Fermi-LAT:2015kyq}. In  \cite{Fermi-LAT:2015kyq} three dark matter density profiles are investigated. An NFW dark matter density distribution with scale radius $r_s = 20\, \mathrm{kpc}$ normalized to a local dark matter density of $0.4\,\mathrm{GeV/cm}^3$ at the Earth's location, a contracted NFW profile with $\gamma=1.3$, and an Einasto profile. However, in our work, we will use the updated results from \cite{Foster:2022nva} which used 14 years of data assuming an NFW and a contracted NFW-like profile with $\gamma = 1.25$ as suggested in the dark matter interpretation of the GeV Galactic Center excess \cite{Daylan:2014rsa,Fermi-LAT:2015sau}. This contracted dark matter density profile emerges from simulations that include baryonic effects in the formation of galaxies ~\cite{Hopkins:2017ycn,Dessert:2022evk}. In the end, one finds a steeper dark matter density profile for the Galactic Center. In Fig. \ref{fig:prof} we plot the density as a function of the distance to the center of our galaxy for the different density profiles adopted in this work.

Besides the Fermi-LAT data, we will also use H.E.S.S. data in the direction of the galactic center. H.E.S.S. collaboration has accumulated 10 years of observations of the galactic center which represents 254 hours of exposure. In this analysis, H.E.S.S. adopted the usual ON and OFF regions to improve the signal over the background ratio. In the end, H.E.S.S. reported a limit of $\left< \sigma v \right> < 10^{-30} cm^3 s^{-1}$ for a $1$~TeV dark matter mass.
  
With those limits at hand, we reinterpret them in terms of effective operators for both scalar and fermion dark matter. In the next section, we describe the effective field theory (EFT) approach used.

\section{Effective Field Theory}
\label{sec:ops}

As we are not considering a full model and we aim to describe the dark matter annihilation into gamma-rays using EFT, we assume that the dark matter particle, $\chi$, to be stable via some discrete symmetry, and a singlet under the SM gauge group.  We are interested in operators containing at least one photon. In principle, we expect the higher dimension operators to be subdominant as they involve additional heavy particles. One should keep in mind that there are pros and cons regarding the EFT approach. It is valid for low momentum transfer. In other words, momentum transfer is much smaller than the mediator mass that connects the dark sector and the visible one. For non-relativistic dark matter annihilation, the momentum transfer is of order to the dark matter mass. Nevertheless, the EFT represents an easy way to test dark matter models. In Table~\ref{Operatorlist-scalar} and Table~\ref{Operatorlist-fermion} we present the lowest-order effective operators that generate gamma-ray lines for scalar and fermion dark matter, respectively.

Our effective field theory is constructed as the Standard Model, plus a dark matter particle $\chi$ which we allow to be either
a complex scalar or Dirac fermion.  In case the dark matter particle is its own antiparticle, the final result will change by a factor of two. We have expressed our EFT in terms of the symmetry field, where $W_{\mu\nu}$ and $B_{\mu\nu}$ are the field strength tensors of the $SU(2)_L$ and $U(1)_Y$ groups, respectively. Consequently in Tables~\ref{Operatorlist-scalar}-\ref{Operatorlist-fermion}, the effective operators are written in terms of the symmetry fields, which after spontaneous symmetry breaking are connected to the mass eigenstates,

\begin{align}
B_\mu=&A_\mu\cos\theta_W-Z_\mu\sin\theta_W\\\nonumber
W^3_\mu=&A_\mu\sin\theta_W+Z_\mu\cos\theta_W
\end{align}where $A_\mu$ and $Z_\mu$ are the photon and $Z$ boson fields respectively, and $\theta_W$ is Weinberg mixing angle.

Up to dimension six, these effective operators represent the full list of possible annihilation channels into gamma-ray lines, and for this reason, our analysis will be restricted to them.

\begin{figure}
\begin{subfigure}{.5\textwidth}
  \centering
  \includegraphics[width=0.95\linewidth]{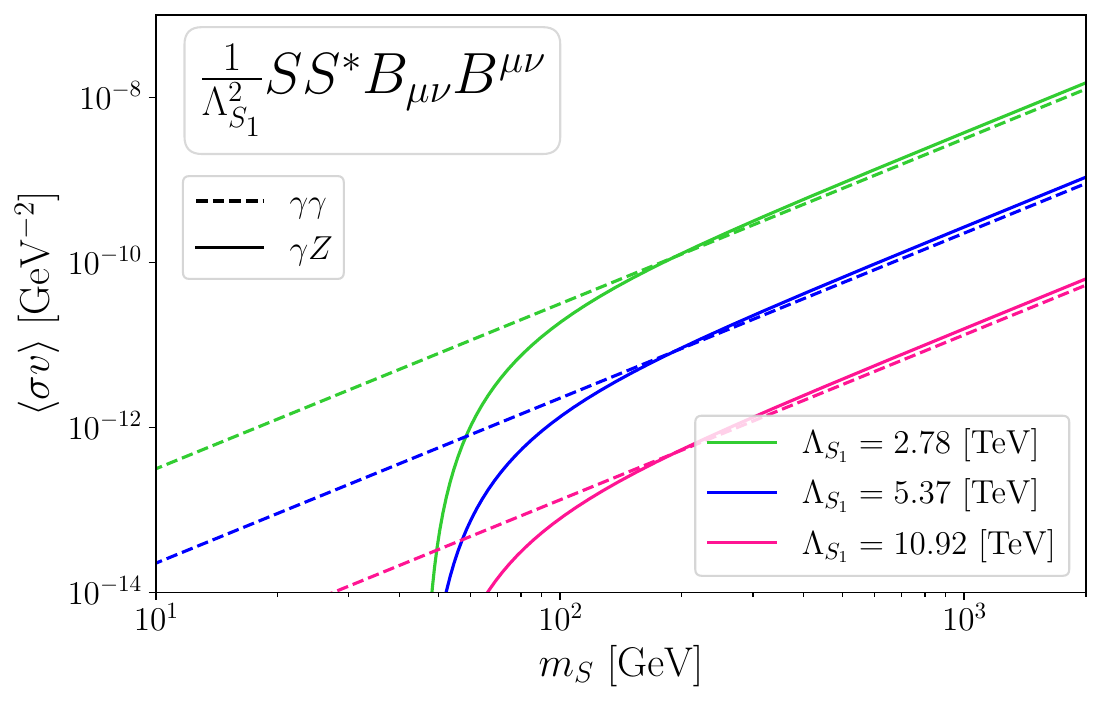}
  \label{fig:sv-S1}
\end{subfigure}
\begin{subfigure}{.5\textwidth}
  \centering
  \includegraphics[width=0.95\linewidth]{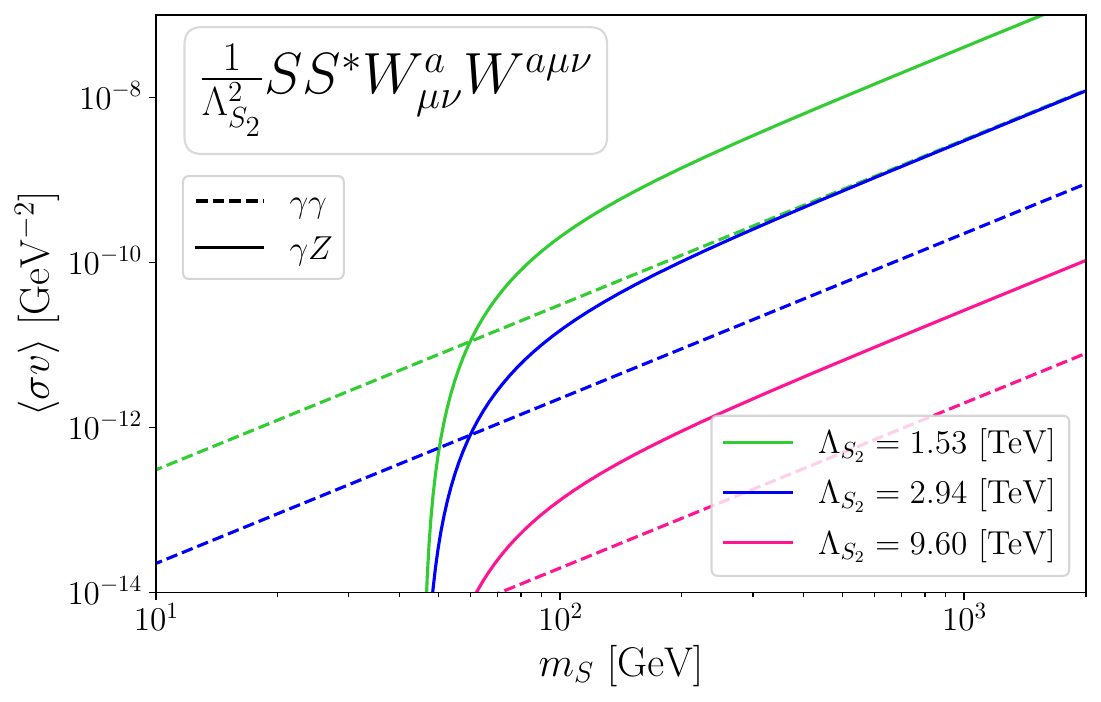}
  \label{fig:sv-S2}
\end{subfigure} \\
\begin{subfigure}{\textwidth}
  \centering
  \includegraphics[width=.45125\linewidth]{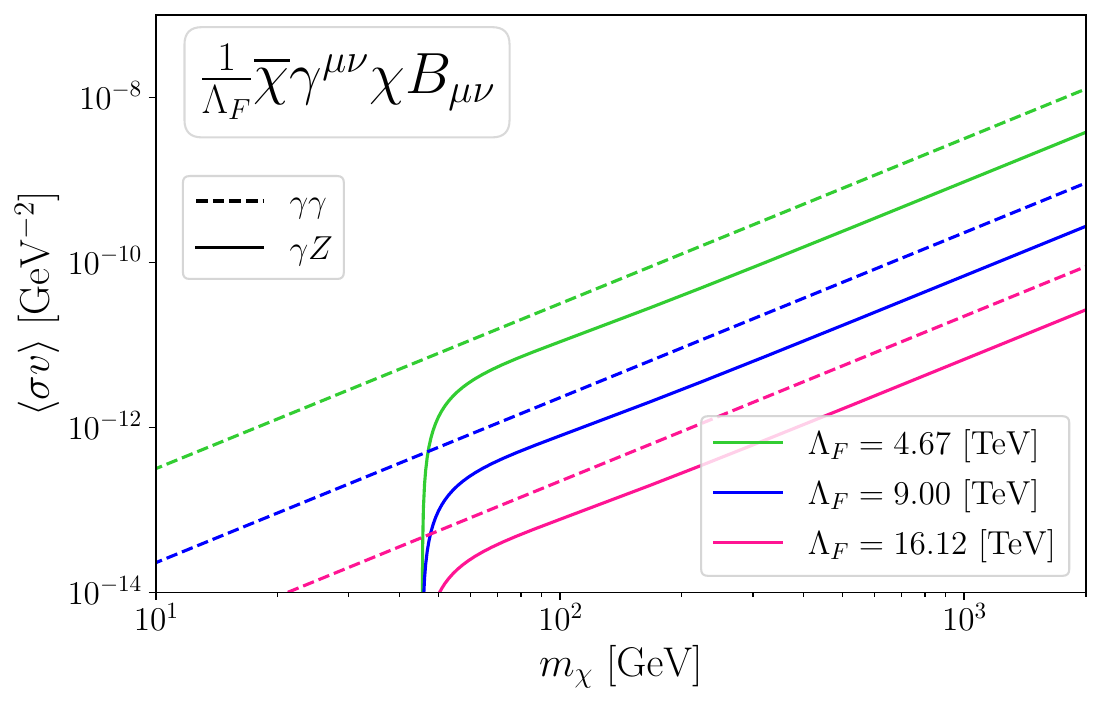}
  \label{fig:sv-F}
\end{subfigure}
\caption{Thermally averaged annihilation cross section for scalar (S) and fermion ($\chi$) dark matter annihilating into $\gamma \gamma$ or $\gamma Z$ for the lowest dimensional operators.}
\label{sigma-v}
\end{figure}

\section{Cross Sections}

Dark matter particles should be cold today. In the galactic halo, the velocity dispersion is estimated to be $v \sim 10^{-3}c$. Hence, we can express the dark matter annihilation cross section as a series in terms of the velocity. Denoting $p_1, p_2$ to be the incoming dark matter particle momenta, $p_3$ and $p_4$ will be the photon momenta. In the case of the $\gamma Z$ line, the Z boson momentum will be represented by $p_4$.

The differential cross section is written
\begin{equation}
\frac{d\sigma}{d\Omega}= \frac{E_3}{256\pi^2~E^3~v} \overline{|{\cal M} |^2}
\label{eqCS}
\end{equation}
where $E = m_\chi + {\cal O}(v^2)$ is the energy of each dark matter particle, $v$ is the dark matter velocity, and $E_3 = |\vec{p}_3|$ is the energy of the outgoing line photon, $|{\cal M} |^2$ is the matrix element ${\cal M}$ averaged over
initial dark matter spins (if any) and summed over final state particle spins.

\begin{center}
\begin{table}
\centering
\begin{tabular}{|c|c|c|c|c|}
\hline
Fermi-LAT& $m_{S/\chi}$ [GeV] & $\frac{1}{\Lambda^2} SS^*B_{\mu \nu}B^{\mu \nu}$& $\frac{1}{\Lambda^2} SS^*W^a_{\mu \nu}W^{a\mu \nu}$ & 	$\frac{1}{\Lambda} \bar{\chi} \gamma^{\mu \nu} \chi B_{\mu \nu}$ \\
\hline
& 10 & 2.78 & 1.53 & 4.67  \\
$\Lambda_{\rm min}$ [TeV] & 100 & 5.37 & 2.94 & 9.00  \\
& 1000 & 10.92 & 9.60 & 16.12  \\
\hline
\hline
H.E.S.S.& $m_{S/\chi}$ [GeV] & $\frac{1}{\Lambda^2} SS^*B_{\mu \nu}B^{\mu \nu}$& $\frac{1}{\Lambda^2} SS^*W^a_{\mu \nu}W^{a\mu \nu}$ & 	$\frac{1}{\Lambda} \bar{\chi} \gamma^{\mu \nu} \chi B_{\mu \nu}$ \\
\hline
& 300 & 6.38 & 5.59 & 9.50  \\
$\Lambda_{\rm min}$ [TeV] & 1000 & 12.10 & 10.63 & 17.84  \\
& 10000 & 18.89 & 16.60 & 27.96  \\
\hline

\end{tabular}
\caption{Lower bounds on the effective energy scale $\Lambda$ using Fermi-LAT (NFWc) and H.E.S.S. (Einasto) data on gamma-ray lines.}
\label{benchmarks}
\end{table}
\end{center}

\subsection{Scalar Dark Matter}

Operators S1-S4 have the form
$\chi \chi^\ast F^{\mu\nu} F_{\mu\nu}$ or $\chi \chi^\ast  F^{\mu\nu} \widetilde{F}_{\mu\nu}$, where $F^{\mu \nu} = B^{\mu \nu}$ or
$W^{\mu \nu}$, and $\chi = S$. For the operators that involve $F^{\mu\nu} F_{\mu\nu}$, where $F^{\mu \nu} = B^{\mu \nu}$ or $W^{\mu \nu}$, the matrix element for $SS^* \rightarrow \gamma \gamma$ reads,
\begin{equation}
\mathcal{M}_{SS^* \rightarrow \gamma \gamma} = 2 Y \left[ (p_3\cdot p_4)(\varepsilon_3 \cdot \varepsilon_4)- (p_3\cdot \varepsilon_4)(p_4 \cdot \varepsilon_3) \right],
\end{equation}
where $p_i$ is the momentum of the outgoing photon with polarization vector $\varepsilon_i$, $Y \equiv \cos^2 \theta_W /\Lambda$, and $\theta_W$ is Weinberg angle. As a result, we get the following spin-averaged matrix element squared,
\begin{equation}
\sum_{\varepsilon_e, \varepsilon_4} |\mathcal{M}_{SS^* \rightarrow \gamma \gamma }|^2 = 4YY^\dagger \frac{s^2}{2} = \frac{2 s^2 \cos^4\theta_W}{\Lambda^4},
\end{equation}
where $s=(p_3+p_4)^2$.

For the $SS^\ast \rightarrow \gamma Z$, where $p_4$ is the momentum of the Z boson we get,  
\begin{equation}
\sum_{\varepsilon_e, \varepsilon_4} |\mathcal{M}_{SS^* \rightarrow \gamma Z}|^2 = 4YY^\dagger \frac{1}{2} \left( s-m_Z^2 \right)^2 = \frac{2  \sin^2 (2 \theta_W)}{\Lambda^4} \left( s- m_Z^2 \right)^2.
\end{equation}
Applying these results to Eq.\eqref{eqCS}, the average annihilation cross sections for the S1 (S3) operator into $\gamma\gamma$ and $\gamma Z$ lines are, 
\begin{eqnarray}
\left< \sigma_{S S^* \rightarrow \gamma \gamma} \, v_{\rm rel}\right> &=& \frac{1}{\pi} \cos^4(\theta_W) \frac{m_S^2}{\Lambda_{S1}^4}, \label{eq:sv1gg} \\
\left< \sigma_{S S^* \rightarrow \gamma Z} \, v_{\rm rel}\right> &=& \frac{1}{\pi} \sin^2 (2\theta_W) \frac{m_S^2}{\Lambda_{S1}^4} \left( 1 - \frac{m_Z^2}{4m_S^2}  \right)^3,
\end{eqnarray}
and for the S2 (S4) operator are,
\begin{eqnarray}
\left< \sigma_{S S^* \rightarrow \gamma \gamma} \, v_{\rm rel}\right> &=& \frac{1}{\pi} \sin^4(\theta_W) \frac{m_S^2}{\Lambda_{S2}^4}, \\
\left< \sigma_{S S^* \rightarrow \gamma Z} \, v_{\rm rel}\right> &=& \frac{1}{\pi} \sin^2 (2\theta_W) \frac{m_S^2}{\Lambda_{S2}^4} \left( 1 - \frac{m_Z^2}{4m_S^2}  \right)^3.
\label{Annscalar}
\end{eqnarray}

\begin{figure}[h!]
\begin{subfigure}{.5\textwidth}
  \centering
  \includegraphics[width=\linewidth]{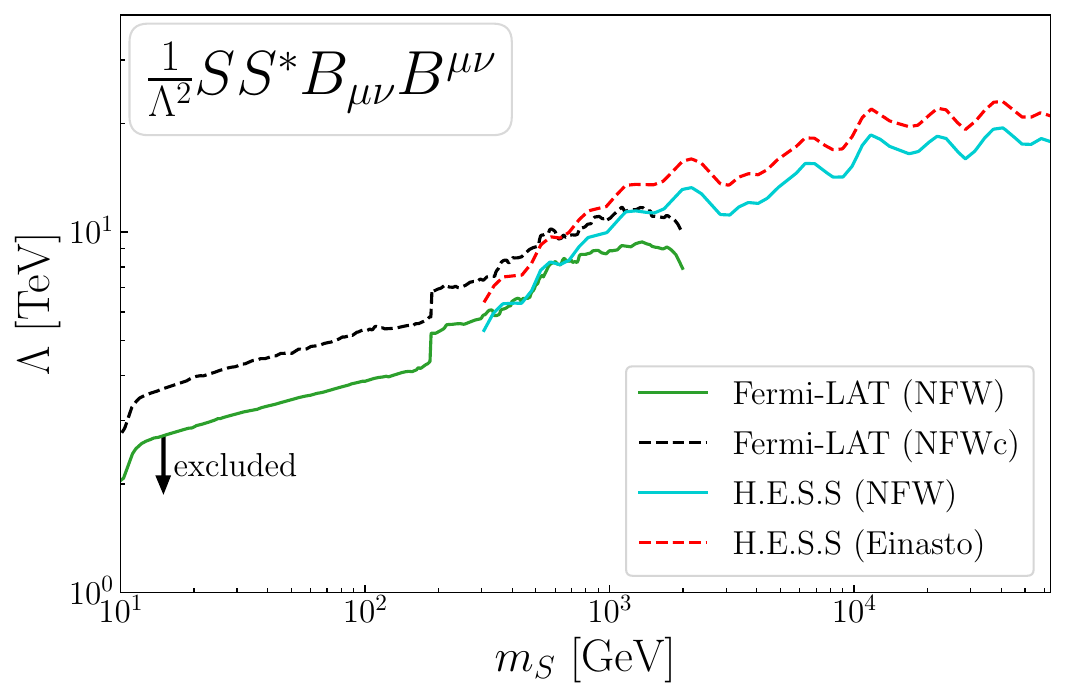}
\label{fig:sfig1}
\end{subfigure}
\begin{subfigure}{.5\textwidth}
  \centering
  \includegraphics[width=\linewidth]{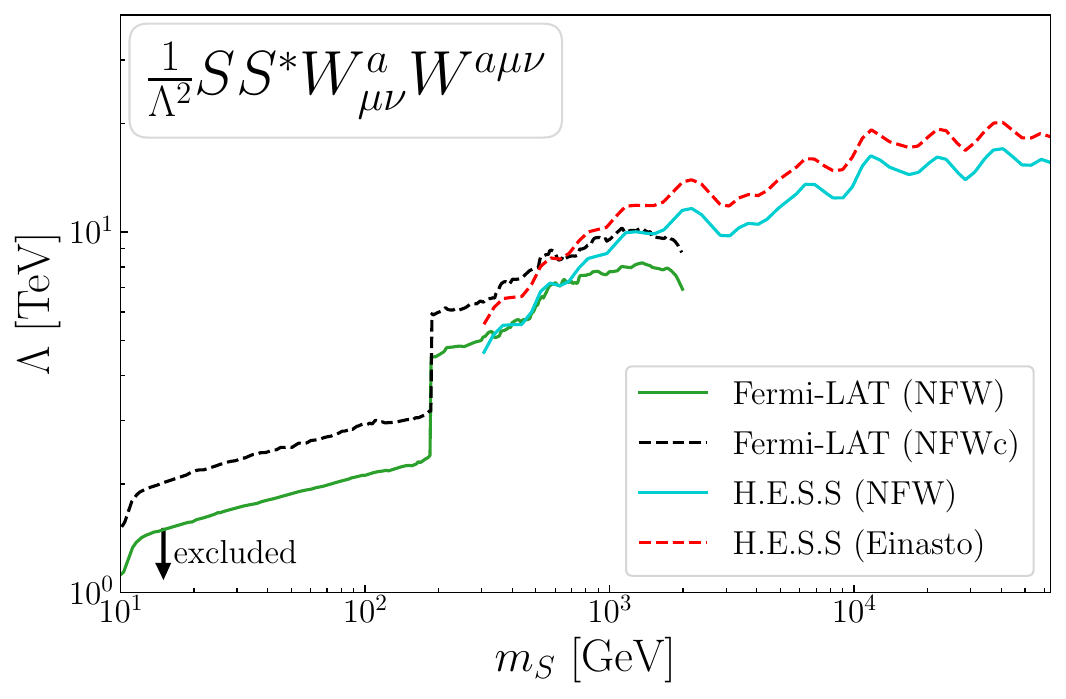}
 \label{fig:sfig2}
\end{subfigure} \\
\begin{subfigure}{\textwidth}
  \centering
  \includegraphics[width=.5\linewidth]{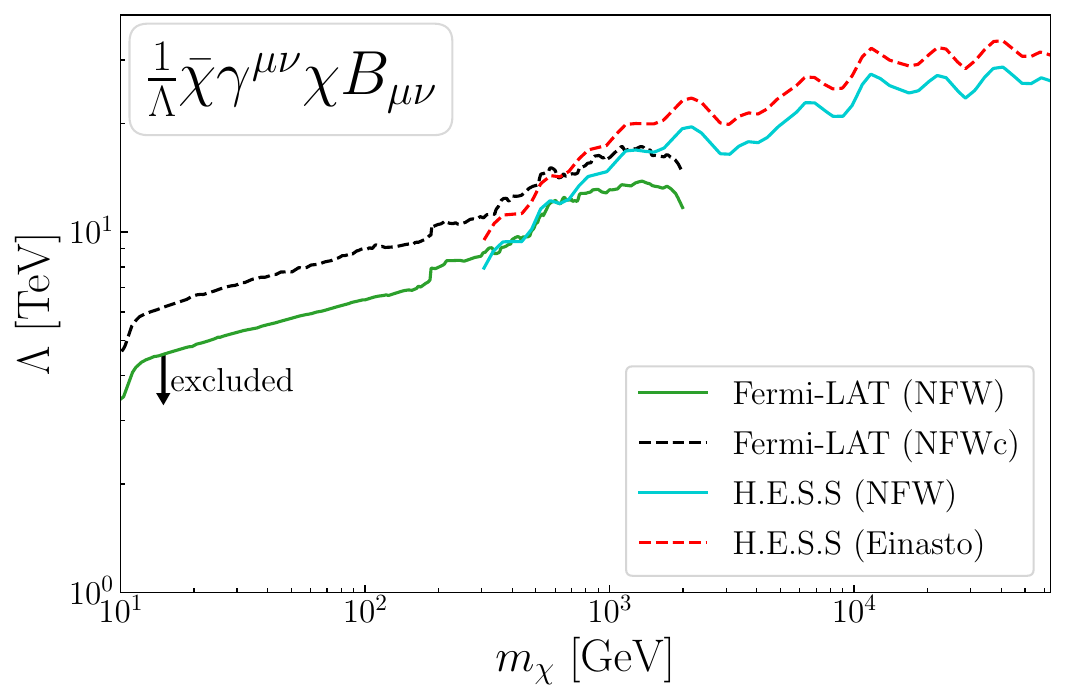}
 \label{fig:sfig2}
\end{subfigure}
\caption{In green, lower bounds for the energy scale $\Lambda$ with respect to the dark matter mass $m_\chi$. The black dashed line shows the lower limits when the DM halo profile scales as $\rho \propto r^{-1.25}$ which is favored by the DM-annihilation interpretation of the galactic center excess. In red and cyan, we show the lower limits from H.E.S.S. 10 years of observations at the inner part of the Milky Way halo using an Einasto and NFW dark matter density profile, respectively.}
\label{results}
\end{figure}

\subsection{Fermionic Dark Matter}

In a similar vein, for operators F1 and F2, we get,
\begin{eqnarray}
\left< \sigma_{\chi\bar{\chi} \rightarrow \gamma \gamma} \,v_{\rm rel} \right> &=& \frac{8}{\pi} \cos^4(\theta_W) \frac{m_\chi^2}{\Lambda_F^4}, \\
\left< \sigma_{\chi\bar{\chi} \rightarrow \gamma Z} \,v_{\rm rel} \right> &=& \frac{2}{\pi} \sin^2(2\theta_W) \frac{m_\chi^2}{\Lambda_F^4} \left(1+\frac{m_Z^2}{4m_\chi^2} \right)^2  \left( 1-\frac{m_Z^2}{4m_\chi^2} \right) , \label{Annfermion}
\end{eqnarray}
for either operator. With the expressions for the annihilation cross sections, we can now place a limit on the effective energy scale $\Lambda$ for each operator, as we do in the next section. 

We plot in Fig.\ref{sigma-v} the results for the average dark matter annihilation cross-section for scalar and fermionic dark matter as a function of the dark matter mass for different values of $\Lambda$. In the upper panels, dedicated to scalar dark matter annihilating into $\gamma \gamma$ and $\gamma Z$ lines through the $S S^* B_{\mu \nu}B^{\mu \nu}$, $S S^* W_
{\mu \nu} W^{\mu \nu}$, we observe that the annihilation into $B_{\mu \nu}B^{\mu \nu}$ yields a larger annihilation cross section into $\gamma\gamma$. For instance, taking $m_S=10$~GeV $\left< \sigma v \right>=10^{-12}$~GeV for both operators if the effective energy scale of the operator $S S^* B_{mu \nu}B^{mu \nu}$ is twice as large. This has to do with the symmetry factor that appears due to the presence of identical particles in the final state. The drop in the annihilation cross section occurs due to the term proportional to $1- m_Z^2/(4m_\chi^2)$ taking into account the energy resolution of the experiments that we will address below. In the bottom panel of Fig.\ref{sigma-v}, we focus on the fermionic case, the average annihilation cross section is larger than the scalar case. One can see that by comparing Eq.\eqref{Annscalar} and Eq.\eqref{Annfermion}. Therefore, we have taken $\Lambda_F=3.4-13$~TeV to keep $\left< \sigma v\right> \simeq 10^{-12}$~GeV, for $m_\chi=10$~GeV. 
\section{Results}

As aforementioned, we use recent searches for gamma-ray lines from dark matter annihilation using Fermi-LAT and H.E.S.S. data to constrain dark sectors through an EFT approach. As we have shown in the previous section, an effective operator can give rise to  $\gamma\gamma$ and $\gamma Z$ lines. The $\gamma Z$ lines happen as long as $m_\chi > m_Z/2$. Therefore, in general we need to account for both $\gamma \gamma$ and $\gamma Z$ lines thus,
\begin{equation}
\left< \sigma \, v_{\rm rel} \right>_{\rm ann} = \left< \sigma_{XX \rightarrow \gamma \gamma} \,v_{\rm rel} \right> + \left< \sigma_{XX \rightarrow \gamma Z} \,v_{\rm rel} \right>   \Theta \left[ \frac{\Delta E}{E} - \frac{m_Z^2}{4m_X^2} \right],
\label{Eqsv}
\end{equation} where $X$ represents a complex scalar or a Dirac fermion, $\Delta E/E$ is the energy-dependent Fermi-LAT energy resolution, and $\Theta$ is the Heaviside step function.  The telescopes do not measure the energy of the photons with infinite precision. The finite energy resolution of the telescope, also known as energy dispersion has been encoded in the instrument response function. For Fermi-LAT, the energy resolution is <10\% between 1 GeV and 100 GeV, which is sufficient to limit the spectral distortion to less than 5\% in this energy range \cite{Fermi-LAT:2013jgq}. We will assume it to be 5\% throughout.  The energy resolution of HESS is poorer, reading 10\% \cite{HESS:2018cbt}. Eq.\eqref{Eqsv} accounts for the fact that the final state photons have different energies given by,
\begin{eqnarray}
E_{\gamma \gamma} & = & m_X, \\
E_{\gamma Z} & = & m_X - \frac{m_Z^2}{4m_X}, 
\end{eqnarray}
and when their relative difference $(E_{\gamma \gamma}-E_{\gamma Z})/E_{\gamma \gamma}$ is smaller than the energy resolution of Fermi-LAT, then the detector cannot tell the difference between the two possible final states above.  We used the energy resolution of each instrument to derive the correct bound on $\left< \sigma v \right>$ using Eq.\eqref{Eqsv}.

That said, our results are summarized in Fig. \ref{results}. In Fig. \ref{results} we plot the Fermi-LAT limits adopting an NFW profile and NFWc profile. As discussed previously, the NFWc density profile is more cuspy than the NFW. For this reason, the constraint on the effective energy scale that controls the dark matter annihilation cross section is stronger.  We also plotted the limits from H.E.S.S. using an NFW and an Einasto profile. The importance of having both instruments is clear in Fig.\ref{results}. Regardless of the effective operator in question, H.E.S.S. yields the most stringent limits on the effective energy scale for dark matter masses above 1 TeV.  In particular, from the leftmost upper panel, we conclude that the effective annihilation of a scalar dark matter particle annihilating into $B_{\mu\nu} B^{\mu\nu}$ is excluded up to energy scales of $10$~TeV for $m_S=1$~TeV, and up to $20$~TeV for $m_S \simeq 600$~TeV. H.E.S.S. loses sensitivity for dark matter masses below $300$~GeV because H.E.S.S. telescopes are designed to probe very high energy gamma-rays, conversely to Fermi-LAT. Fermi-LAT is the best instrument to probe dark matter particles in the mass range between $10$~GeV - $300$~GeV. For the fermionic case, the bound on the effective energy scale is significantly stronger because the annihilation cross section into gamma-ray lines is larger. For instance, with an NFW profile, Fermi-LAT imposes at most $\Lambda > 2$~TeV for $m_S=10$~GeV, whereas $\Lambda > 3.5$~TeV for $m_\chi=10$~GeV. Regardless of the dark matter nature, we notice an interesting complementarity between Fermi-LAT and H.E.S.S. for dark matter masses between $300$~GeV and $1$~TeV. We have explicitly summarized the strongest limits from Fermi-LAT and H.E.S.S. on the effective energy scale in Table \ref{benchmarks} which arise from adopting either an NFWc or Einasto density profile.

\section{Conclusions}

We have used 14 years of Fermi-LAT and 10 years of H.E.S.S. observations in the direction of the Galactic Center to constrain the emission of gamma-ray lines from scalar and fermionic dark matter taking into account the energy resolution of the instruments. We have described the production of $\gamma\gamma$ and $\gamma Z$ lines in terms of the lowest-order effective operators, which are normalized by an effective energy scale ($\Lambda$). Considering different dark matter density profiles such as NFW, NFWc, and Einasto we exhibit the Fermi-LAT and H.E.S.S. limits on $\Lambda$ as a function of the dark matter mass. An NFWc density profile which is favored by the GeV excess observed in the Galactic center yielded stronger limits for featuring a more cuspy profile. Fermi-LAT turned out to be the best probe for dark matter masses below $300$~GeV, with H.E.S.S. being more restrictive to gamma-ray lines all the way up to $600$~TeV. The telescopes share similar sensitivities in the $300$~GeV - $1$~TeV mass range. In particular, for both scalar and fermion dark matter, Fermi-LAT imposes $\Lambda > 10$~TeV for a dark matter mass of $1$~TeV, whereas H.E.S.S. places a lower limit $\Lambda > 20$~TeV, for a $10$~TeV dark matter mass.

\acknowledgments

The authors thank Jacinto Paulo for the discussions. This work was supported by Simons Foundation (Award Number:1023171-RC), FAPESP Grant 2018/25225-9, 
2021/01089-1, 2023/01197-4, ICTP-SAIFR FAPESP Grants 2021/14335-0, CNPq Grant 307130/2021-5, and ANID-Millennium Science Initiative Program ICN2019\textunderscore044. The authors acknowledge the National Laboratory for Scientific Computing (LNCC/MCTI, Brazil) for providing HPC resources of the SDumont supercomputer (\url{http://sdumont.lncc.br}). GG is supported by grant 2022/16580-5, S\~{a}o Paulo Research Foundation (FAPESP). LA acknowledges the support from Coordenação de Aperfeiçoamento de Pessoal de Nível Superior (CAPES) under grant 88887.827404/2023-00. L.G. thanks the support from Coordenação de Aperfeiçoamento de Pessoal de Nível Superior (CAPES) under grant No. 88887.704425/2022-00. VdS acknowledges CNPq grant No. 303942/2019-3.

\bibliographystyle{JHEPfixed}
\bibliography{references}

\end{document}